% version of 1.10.01 

\input harvmac 

%%%%%%%%%%%%%%%%%%%%%%%%%%%%%%%%%%%%%%%%%%%%%%%%%%%%%%%%%%%%%%%%
\input epsf.tex
%%%%%%%%%%%%%%%%%%%%%% macros for figures %%%%%%%%%%%%%%%%%%%%%%%
\newcount\figno
\figno=0
\def\fig#1#2#3{
\par\begingroup\parindent=0pt\leftskip=1cm\rightskip=1cm\parindent=0pt
\baselineskip=11pt
\global\advance\figno by 1
\midinsert
\epsfxsize=#3
\centerline{\epsfbox{#2}}
\vskip 12pt
\centerline{{\bf Fig. \the\figno}} 
#1\par
\endinsert\endgroup\par}
\def\figlabel#1{\xdef#1{\the\figno}}
\def\encadremath#1{\vbox{\hrule\hbox{\vrule\kern8pt\vbox{\kern8pt
\hbox{$\displaystyle #1$}\kern8pt}
\kern8pt\vrule}\hrule}}

%%%%%%%%%%%%%%%%%%%DEFINITIONS%%%%%%%%%%%%%%%%%%%%%%%%%%%%%%%%%

\def\blank#1{}
\def\za{\alpha} \def\gga{\overline{\gamma}} \def\ga{\gamma}
 \def\lla{\overline{\Lambda}}

%%%%%%%%%%%%%%%%%%%%%
\def\za{\alpha}  \def\zg{\gamma} 
   
\def\zk{\kappa} \def\zl{\lambda}

\def\bL{\bar{\Lambda}}
\def\zG{\Gamma}   
\def\zL{\Lambda}  

\def\IZ{Z\!\!\!Z}
\def\[{\,[\!\!\![\,} \def\]{\,]\!\!\!]\,}
\def\dC{C\kern-6.5pt I}

\def\by{\bar y}

\def\bW{\overline W}
        \def\CC{{\cal C}}
        
\def\CG{{\cal G}}

        \def\CR{{\cal R}}
        \def\CU{{\cal U}}

\def\un{{\bf 1}}
\def\ochi{\overline{\chi}}

%xxxxxxxxxxxxxxxxxxxxxxxxxxxxxxxxxxxxx

%%%%%%%%%%%%%%%%%%%%%%% amsTEX characters %%%%%%%%%%%%%%%%%%

\def\IC{\relax\hbox{$\inbar\kern-.3em{\rm C}$}}

%\def\msy{y }
%\message{Do you have the AMS fonts (y/n) ?}\read-1 to 
%\msan\ifx\msan\msy
%\input mssymb
\input amssym.def
\input amssym.tex
\def\IZ{\Bbb Z}\def\IC{\Bbb C}
\def\gg{\goth g} 
%\else
%\def\gg{g}
%\fi

%%%%%%%%%%%%%%%%%%%%%%

%%%%%%%%%%%%%%%%%%%%%%%%%%%%%%%%%%%%%%%%%%%%%%%%%%%%%%%%%%%%%%%%%
%
%           Current Definitions
%

 \def\bgo{\overline{\goth g}} %%% (affine) Lie
        %%% Cartans

                   %%% (simple/real) roots
\def\Rr{\Delta^{\rm re}} 
                      % horizontal Weyl vector
                      % horizontal roots 

                       %%% weight/root lattice
 
\def\FW{\Lambda}\def\fw{\overline\FW}       % (horizontal)
                                            %fund.weights 

\def\bW{\overline W} \def\tW{\tilde W}      %%% Weyl groups

                 % KW action on left
                       % horiz. KW group
                                 % group unit
\def\UNIT{\blacktriangleleft\kern-.6em\blacktriangleright}

                               %%% translations

\def\ts#1,#2{{\tt e}^{#1\zk#2}}
%\def\ts#1,#2{\tr_{#1#2}}

                %%% cyclic groups  

%\def\gg{\gamma} \def\tg{\tilde\gamma}

\def\iii{\iota}                           %%% the map `iota'

\def\CC{{\cal W}^{(+)}} \def\tC{\tilde\CC} %%% chambers

\def\UU{{\cal U}}                         %%% `fiber' over w_0

                          %%% supports
         %%% `hats' or `head tiles' 
   %#1#2{{\cal V}_{#1}^{(#2)}}   %%% supp of `Verma'
                 %%% supp of f.d.  module
\def\GG{{\cal G}}                 %%% supp of f.d. `module'
                          %%% ordinary Verma module
                %%% ordinary Verma module 
                                             %      of type #2

\def\cc{\chi} \def\bc{\overline\cc}       %%% characters

                                 %%% `horizontal' denominator
                           %%% `invariant' denominator
                        %%% generalized `denominator'
                                 % Kostant partition

\def\ml{m}                                 %%% multiplicities
\def\bm{\overline{m}}                 %   ordinary multiplicities
                               %%% fundamental characters

\def\CR{{\goth W}} 
                    %%% rings 

  %%% filtrations
  %%% filtrations
  %%% filtrations
  %%% filtrations

            %%% number rings, cones, etc.
\def\IZp{\IZ_{\ge 0}} 

\def\IQ{{\Bbb Q}}

  %%% triality

                          %%% level

        %%% dimensions

                    %\def\hk{{\hat\kmath}}

\def\ol#1{\overline{#1}}

\def\PROP{\medskip\noindent{\bf Proposition:}\ \ }

\def\LEMMA{\medskip\noindent{\bf Lemma:}\ \ }

\def\endPROOF{\quad$\square$\medskip}
%\def\endPROOF{\quad$\blacksquare$\medskip}
%\def\SECTION#1{\bigskip\noindent{\bf #1}\medskip}

%%%%%%%%%%%%%%%%%%%%%%%%%%%%%%%%%%%%%%%%%%%%%%%%%%%%%%%%%%%%%%%%%

%AUTHOR'S inputs, e.g., harvmac, amstex, phyzzx, etc, 
%macros, definitions  

\hfuzz=10pt

\font\ttfont=cmbx10 scaled\magstep3 
\font\sfont=cmbx10 scaled\magstep1 
\font\rfont=cmr10 scaled\magstep1 

\vsize=8.5truein \hsize=6truein
%\nopagenumbers 
\topskip=0.5truein
\raggedbottom 
\abovedisplayskip=3mm 
\belowdisplayskip=3mm 
\abovedisplayshortskip=0mm 
\belowdisplayshortskip=2mm 
\normalbaselineskip=12pt 
\normalbaselines

%%%%%%%%%%%%%%%%%%%%

%%%%%%%%%%%%%%%%%%%%%%%%%

\lref\AY{Awata H. and  Yamada Y., 
%Fusion rules for the fractional level $\widehat{sl}(2)$  algebra,
         Mod. Phys. Lett. {\bf A7}, 1185 (1992).} 
		 %--1195 (1992).}

\lref\FMa{Feigin B.L. and Malikov F.G.,
%        Fusion algebra at a rational level and cohomology of 
%           nilpotent subalgebras of $\widehat{sl}(2)$,
      Lett. Math. Phys. {\bf 31}, 315 (1994).}
	  %--325 (1994).}
\lref\FMb{Feigin B.L. and Malikov F.G., 
 Modular functor and representation 
        theory of $\hat{sl(2)}$ at a rational level, q-alg/9511011,
in
         {\it Operads}: Proceed.
%ings
 of Renaissance Conferences,  
         Cont. Math. {\bf 202},  357, eds.  J.-L. Loday et al
% J.D. Stasheff and   A.A. Voronov, eds.
          (AMS, Providence, Rhode Island,  1997).}
% q-alg/9511011.} 

%\lref\MFF{Malikov, F.G., Feigin, B.L., and Fuks, D.B.,
%     	  Funct.\ Anal.\ Pril.\ {\bf 20, 2} 25 (1987) 25.}

\lref\FGP{Furlan P., Ganchev A.Ch. and Petkova V.B., % 1990,
%             Quantum groups and fusion rule multiplicities,
              Nucl. Phys. {\bf B343}, 205
			  %--227 
(1990).} 

\lref\FGPa{Furlan P., Ganchev A.Ch. and Petkova V.B.,
% Fusion rules for admissible representations of affine algebras:
%           the case of $A_2^{(1)}$,  
         Nucl. Phys. {\bf B518} [PM], 645  (1998).}
		 %--668 (1998).}
		 %, hep-th/9709103.}
 
\lref\FGPb{Furlan P., Ganchev A.Ch. and Petkova V.B.,
%         An extension of the character ring of $sl(3)$ and its quantisation,
		Comm. Math. Phys. {\bf 202 }, 701 (1999).}
		%--733 (1999).}
		%, math.QA/9807106.}
		
\lref\GPW{Ganchev A.Ch., Petkova V.B. and Watts G.M.T., 
%A note on decoupling conditions for generic level $\widehat{sl}(3)_k $
%and fusion rules, 
Nucl. Phys. {\bf B571} [PM], 457 (2000).}
%--478 (2000).}
% hep-th/9906139.}
 
\lref\Hump{Humphreys J.M., 1990, {\it Reflection Groups and
           Coxeter Groups}, (Cambridge University Press, 1990).} 

\lref\K{Kac V.G., {\it Infinite-dimensional Lie Algebras}, third
        edition, (Cambridge University Press, 1990).}

\lref\KW{Kac V.G. and  Wakimoto M.,  
%1988,
%      Modular invariant representations of infinite-dimensional 
%      Lie algebras and superalgebras,
    Proc. Natl. Sci. USA {\bf 85}, 4956 
%\semi
	%--4960 
(1988)   \semi
 Kac V.G. and  Wakimoto M.,
% 1989,
%         Classification of modular invariant 
%                 representations of affine algebras,
               Adv. Ser. Math. Phys. vol {\bf 7},  138  (1989) %--177. 
(World Scientific, Singapore)\semi
%, 1989)         \semi
 Kac V.G. and  Wakimoto M., 1990,
% Branching functions for winding subalgebras and tensor products, 
          Acta Applicandae Math. {\bf 21}, 3
		  %--39 
(1990).}

\lref\Zh{Zhelobenko D.P., 
%1973,
{\it Compact Lie Groups and their Representations}, 
(AMS, Providence, Rhode Island, 1973).}

\lref\MW{Walton M., 
%1990,
%       Fusion rules in Wess-Zumino-Witten models, 
           Nucl. Phys. {\bf B340}, 777
		   %--790 
(1990). }

\lref\Hump{Humphreys J.M., 
%1990, 
{\it Reflection Groups and
           Coxeter Groups}, (Cambridge University Press), 1990).} 

\lref\K{Kac V.G., 
%1990
 {\it Infinite-dimensional Lie Algebras}, third
        edition, (Cambridge University Press,
		 1990).}

\lref\FP{Furlan P. and Petkova V.B., 
% 2000,
Fusion Rings Related to Affine Weyl Groups, hep-th/0007219,
in: Proceed.
%eedings 
of the  International Workshop 
{\it Lie Theory and Its Applications in Physics III}, 
(Clausthal, 1999); eds. H.-D. Doebner et al, 
(World Sci, Singapore,  ISBN 981-02-4421-5) p. 237.}
%-249.}
%hep-th/0007219.}

\lref\GKS{Giveon A.,  Kutasov D. and  Schwimmer  A.,
     Comments on D-branes in AdS$_3$, 
hep-th/0106005.  }
	 
%\lref\admFGP{Furlan P., Ganchev A.Ch. and Petkova V.B., 
%$A_1^{(1)}$ admissible representations --fusion transformations
%and local correlators, Nucl. Phys. {\bf B491}  no. 3 [PM] 635.}
%-658 (1997), hep-th/9608018.}

%\lref\PRY{ J.L. Petersen, J. Rasmussen and M. Yu,  Fusion, Crossing
%      and monodromy in conformal field theory based on $sl(2)$ current
%      algebra with fractional level,  Nucl. Phys. {\bf B 481} 577-624 (1996),
%       hep-th/9607129.}

%%%%%%%%%%%%%%%%%%%%%%%%%%%%%%%%%%%%%%%%%%%%%%%%%%%%%%%%%%%%%%%%%%%%

%%%%%%%%% start of text

\null 
\vskip 1truecm 
 
\centerline{{\ttfont Characters of $\widehat{sl}(4)_k$
fusion algebra}}
\vskip 2truemm
\centerline{{\ttfont  at non-rational level }}
\vskip 3truemm
\vskip 1.5cm

\centerline{{\sfont P. Furlan$^{*,**}$ }
~{\rfont and}~ {\sfont V.B. Petkova$^{\dagger}$}} 

\vskip 0.5 cm 
 
\centerline{
 $^{*}$Dipartimento  di Fisica Teorica
 dell'Universit\`{a} di Trieste and }
\centerline{$^{**}$Istituto Nazionale di Fisica Nucleare
(INFN), Sezione di Trieste,} 
\centerline{Strada Costiera 11,
%\centerline{
34100 Trieste, Italy}

\centerline{$^{\dagger}$Institute for Nuclear Research and
Nuclear Energy,}
\centerline{Tzarigradsko Chaussee 72, 
%\centerline{
1784 Sofia, Bulgaria}
\vskip 1cm 

\centerline{\bf Abstract} 
\vskip 0.2cm

We construct  the fusion ring  of a  quasi-rational
$\widehat{sl}(4)_k\,$ WZNW theory at generic level $k\not \in {\Bbb Q}\,.$ 
It is generated by commutative elements in the group ring
${\Bbb Z}[\tilde{W}]$ of the extended affine Weyl group $\tilde{W}$ 
which extend polynomially  the formal characters of finite dimensional
representations of $sl(4)$.

\noindent
\newsec{ Introduction}
\vskip 0.1cm

The WZNW  models at generic (non-rational) level provide
examples of quasi-rational  conformal field theories (Q-RCFT).
These are theories described by an  infinite discrete spectrum of
representations of the chiral algebra, here
%in our case the affine algebra 
$\gg=\widehat{sl}(n)_k$, and  a  fusion rule  producing a
finite number of terms.  The study of these quasi-rational fusion
rings is motivated by the fact that they  determine, upon
``quantisation'', the fusion rules of the corresponding RCFT,
described by the fractional level admissible representations of $\gg$ \KW.
For the latter there is no sensible  Verlinde formula at hand,
see the two fully worked out $ \widehat{sl}(n)_k\,$ examples so
far, n=2 \AY,\FMa,\FMb\  and n=3  \FGPa,\FGPb,\GPW.  The
quasi-rational fusion rings and their characters are also
important as part of the data of more general CFT with a
continuum spectrum,  on manifolds with or without boundaries;
see, e.g., \GKS\ and  references therein  for the simplest
example of generic level $\widehat{sl}(2)_k$ theory.

%\noindent
Consider the ``preadmissible'' set of representations labelled by
the highest weights
\eqn\preadm{
\eqalign{
\{ 
\overline{\Lambda}  =\bar{y}\cdot
&(\zl'-\zl\, (k+n))\,,  \  k \not \in \IQ \, | \
 \quad  \bar{y}\in \bW\,,\   \zl'\,,\, \zl \in P_+\,,\ 
{\rm s.t.,}\cr &  \langle \zl,\za_i
\rangle \delta + 
\bar{y}(\za_i) \in \Rr_+\,,\,   i=1,2,\dots n-1 \}
}}
where $\bar{y}\cdot \zl$ is the shifted action of the Weyl group $\bW$
of the horizontal algebra $\bgo=sl(n)$, $P_+= \oplus_{i}\ \IZp\
\fw_i\,$ is the 
chamber of integral dominant
weights, $\fw_i$ being the
fundamental weights of $\bgo$,  and $ \Rr_+$ is the set of real
positive roots of $\gg$.  
The ``preintegrable'' subset  ($\bar{y}=\un\,,
\zl=0$,  
$\bL=\zl'\in P_+$) 
has  fusion ring coinciding with the
representation ring of the finite dimensional irreducible representations
(irreps) 
of $\bgo$. 
Its structure constants  
are given by the classical
Weyl - Steinberg formula, which can be ``quantised'' to recover
the fusion rule multiplicities of the integer level integrable
representations \K,\MW,\FGP.  The main ingredient in both the
classical and the ``quantised'' versions of  this
formula is the multiplicity of  states of finite dimensional
irreps of $\bgo$, 
encoded in their formal characters. 
However all these classical data have no direct  
meaning for the second, labelled by non-integer highest weights  of
$\bgo$,  subseries
 of \preadm\ ($\zl'=0$),  we shall deal with below.

Remarkably  
in the simplest $\widehat{sl}(2)_k$ case the %classical
fusion 
characters for the subseries $\zl'=0$ turn out to 
be given by
the formal characters of   finite
dimensional irreps of the  super-algebra  $osp(2|1)$.
Accordingly the quasi-rational fusion ring of the 
$\widehat{sl}(2)_k$ representations \preadm\ coincides with a product of
the representation rings of 
%finite dimensional irreps of
$osp(2|1)$ and $sl(2)$.  Their quantised rational counterpart inherits
this ``hidden'' $\IZ_2$ - graded structure, first noticed  
 in  \FMb.  The group  $\IZ_2$ is 
%isomorphic to 
the
Weyl group $\bW$ of  $sl(2)$ and %although 
the next truly
nontrivial case $n=3$ 
exhibits a 
%kind of a
$\bW$ - graded
algebraic structure too.
The  specific character formulae  established for $n=3$ 
do not extend,  however,  to $n\ge 4$,
that is why it is important to study the simplest next case $n=4$
by methods admitting 
in principle 
a generalisation to arbitrary $n$.

Preliminary results of the present work were announced in
\FP\ and  to make the paper self-contained we repeat in the next
section some of the introductory material there. Our main new
result is the explicit formula in  section 3 for the characters
shown to generate a consistent fusion ring.

\noindent
\newsec{ General setting }

\vskip 0.1cm
 
Consider the subset $\tC$  of the extended affine Weyl group
$\tW=\bW\ltimes t_P= W\rtimes A\,$
\eqn\Ida{
\tC:=\{y\in \tW\,|\, y(\za_i)\in \Rr_+\ \ {\rm for}\ \forall
%\za_i\in \bP
\ i=1,\dots,n-1\}\,.
}
Here    $t_P$ is the subgroup of translations in the weight
lattice $P$ of $\bgo=sl(n)\,,$ 
%$\bW$ and $W$ are the Weyl groups
% of $\bgo$ and $\gg=\widehat{sl}(n)_k$ respectively, 
and $A$ is
the cyclic subgroup of $\tW$, generated by  $\zg= t_{\fw_1}\,
w_1w_2\dots w_{n-1}$, which keeps invariant the set of simple
roots $\Pi=\{\za_0\,,\za_1,\dots, \za_{n-1}\}\,$ of $\gg$.
% $\bP$ is the  set of simple roots of $\bgo$.

Let $k\not \in \IQ\,.$ The subset $\tC$ is a fundamental domain  
%(a ``dominant chamber'') 
in $\tW$ with respect to the right action of $\bW$
\FGPb.  
The set of  weights  $\tC\cdot
k\zL_0 $ (corresponding to the subset $\zl'=0$ of \preadm{},
$y=\bar{y}\,t_{-\zl}$) or,
equivalently, the subset $\tC\subset \tW$ itself, labels the
highest weights $\Lambda$ of maximally 
%(but finitely) 
reducible
Verma modules of $\gg$. Indeed for $\Lambda=y\cdot k\Lambda_0$
and $\beta =y(\za)\,,$ s.t. $y\in \tC$, the Kac-Kazhdan singular
vector criterion holds true for any positive root $\za$ of
$\bgo.\,$ Here $\zL_0$ is the fundamental weight of $\gg$ dual to
the affine root $\za_0\,$ and
%
% and the  shifted action of $\tW$  is given by 
%$\hw\cdot\zL=\hw(\zL+\zr)-\zr\,,$ $\rho $ being the Weyl vector of $\gg$. 
%
the   Kac-Kazhdan reflections are identified with the right
action of $\bW$ on $\tW\,,$ i.e., $w_{y(\za)}\cdot \Lambda = y
w_{\za}\cdot k \Lambda_0\,.$

The factorisation of the submodules generated by these singular
vectors imposes restrictions on the possible three-point
couplings which determine the fusion rules in the CFT. Solving
completely the corresponding null vector decoupling equations is
a difficult problem (see also the Discussion below). Instead we
shall construct recursively a 
fusion ring building
on and extending the approach in \FGPb.  We associate with any
$y\in \tC$ a formal ``character'', an element of the group ring
$\IZ[\tW]\,$ of $\tW$
\eqn\res{
\cc_{y}\,  =  \sum_{z\in\tW\,,\, zy^{-1}\in W} \ml_{z}^{y} \ z\,,
\qquad y\in \tC\,,
}
extended to $\tW$ by
\eqn\resa{
	\cc_{y w}:
 =\det(w)\,\cc_{y}\,, \quad y\in \tC\,, \ w\in
\bW\,.
}
The sum in \res\ is required to run over elements in $\tW$ with
length not exceeding the length of the ``highest weight'' $y$ and
$\ml_{z}^{y}$ are   integer nonnegative multiplicities, yet to be
determined.  
%One introduces the  notion of a generalised ``weight
%diagram'', a finite set, $\GG_{y}=\{z\in \tW \,|\, \ml_{z}^{y}\ne
%0\}\,.$
The characters \res\ are  generalisations of the formal
characters  of finite dimensional irreps
of $\bgo$,
\eqn\class{
\ochi_{\zl}=\sum_{\mu\in \Gamma_{\zl}}\,
\overline{m}^{\zl}_{\mu}\, t_{-\mu} \in  \IZ[t_{-P}]\,, \qquad 
\ochi_{w\cdot\lambda}= \det(w)\, \ochi_{\lambda}\,, \ w\in \bW\,,
 \, \zl\in P_+\,.
}
%
% which are elements in the group ring $\IZ[t_{-P}]$.
The finite set $\GG_{y}=\{z\in \tW \,|\, \ml_{z}^{y}\ne
0\}\,$ generalises the weight diagram $\Gamma_{\zl}$.
%Having
With   a notion of a generalised weight diagram, consider a
formula for the fusion rule multiplicities, generalising the
classical Weyl - Steinberg formula for 
%the characters 
$\ochi_{\zl}$
\eqn\ws{
     \cc_x\ \cc_y 
 = \sum_{z\in {\cal G}_x} \ml^{x}_{z} \ \cc_{zy} 
  = \sum_{z\in \tC} N^{z}_{x,y} \ \cc_{z} \,,
}
\eqn\wsa{
N^{z}_{x,y}
 = \sum_{w\in\bW} \, \det(w)\ \ml^{x}_{z w y^{-1}} \,.  
}
The second equality in \ws\ with the multiplicities given in
\wsa\ is derived as for the usual $sl(n)$ characters, using the
symmetry in \resa\ and the fact that $\tC$ is a fundamental
domain in $\tW$.  
%In particular to the elements of $A$ correspond
% "simple currents", $\cc_a =a\,, a\in A\,,$
%\eqn\ainv{ \cc_a\ \cc_y =\cc_{ay}\,.}%

Introduce a map $ \iii$ of $\tW$  into the root
lattice $Q$  of $\bgo$  \FGPa,\FGPb\
\eqn\mi{
 \iii: \tW\ni y=\by t_{-\zl} \  \mapsto \
     n\,\zl + \by^{-1}\cdot 0 	 \in Q\,.
}
It has  the properties
\eqn\tlg{\eqalign{
\iii(x y ) & = \ol{y}^{-1}(\iii(x)) + \iii(y) \,,\cr
 \iii(y w) & =
 w^{-1} \cdot\iii(y)\,, 
\quad w  \in\bW\,, \cr
 \iii(a x) & =
 \iii(x)\,, \quad a\in A\,,
}}
and the set $\tC$ is expressed alternatively as
$\tC=\{y\in \tW\, |\, \iii(y) \in P_+\}\,.$

In the  $n=3$ case the  coefficients $ \ml_{z}^{y}$ in \res\ are given
as
\eqn\mt{
\ml_{z}^{y}
=\bm^{\iii(y)}_{\iii(z)}\,,
}
$\bm^{\iii(y)}_{\iii(z)}$ being as in \class\ the 
%standard 
multiplicity of the
weight $\mu=\iii(z)$ of the 
%finite dimensional 
representation of
$sl(3)$ of highest weight $\lambda=\iii(y)$.  Similarly the
fusion coefficients \wsa\ are expressed in terms of the structure
constants $\overline{N}^{\iii(z)}_{\iii(x)\,\iii(y)}$ of the $sl(3)$
character ring
\eqn\strci{
N^{z}_{x,y} = \overline{N}^{\iii(z)}_{\iii(x)\,\iii(y)} \,.
}
The generalised weight diagrams ${\cal G}_y$ are 
determined by \mt\ and thus have the structure of the  weight
diagrams $\zG_{\iii(y)}$ of triality zero $sl(3)$
representations, with multiplicities preserved, but with the
weights $\mu\not\in\ $Im$(\iii)$ excluded. The same type of
formulae hold in the simpler $sl(2)$ case where $|{\cal
G}_y|=|\Gamma_{\iota(y)}|$.

However as discussed in \FP\ the definition of the generalised
weight diagram based on \mt,   and hence \strci, has to be
modified in the higher rank cases, since it is not consistent
with the Weyl - Steinberg formula \ws\ in general. The
multiplicities in \res\ are only restricted by the  inequality
$\ml_{z}^{y} \le\bm^{\iii(y)}_{\iii(z)}\,$.

\noindent
\newsec{Fusion character ring} %The case of}  
\vskip 0.1cm

% The cyclic subgroup $A$ of $\tW$ is generated by 
% $\zg= t_{\fw_1}\, \bar{\zg}$, where $ \bar{\zg}=w_1w_2\dots w_{n-1}$
%  is a Coxeter element in $\bW$ generating  the  cyclic subgroup
% $\bar{A}$ of $\bW$. 
% The  Cayley graph of $\tW$  is a  n-sheeted covering
%of the Cayley graph of $W$, see a drawing in \FP\ for $n=4$.

We denote by  $\overline{\CR}$  the character ring of finite
dimensional irreps of $sl(n)$ generated by the formal classical
characters $\ochi_{\zl}$ \class.   They 
%Looked as elements in $\IZ[\tW]$ they
commute with any $w\in \tW$ because of the invariance of the
classical weight multiplicities
$\overline{m}^{\zl}_{w(\mu)}=\overline{m}^{\zl}_{\mu}\,,\,  w\in\bW$.
%We shall also use their invariance up to a sign with respect to
%the shifted action of the Weyl group $\bW$,
%$\ochi_{w\cdot\lambda}= \det(w)\, \ochi_{\lambda}$.

Let $x = t_{-\nu}\overline x \in \tC$. Guided by the $n=2,3$
examples we first introduce the following combinations of
classical $sl(n)$ characters $\ochi_{\zl}$ times powers of the
generator $\zg$ of $A$, parametrised by weights $b\in Q$,
 \eqn\quadr{
 \ochi^{(b)}_x 
= {\rm det}(\overline x) %\{ \ochi_{\nu-b}+
\sum_{i=0}^{n-1}\,
 \ga^i\ \ochi_{\nu- \gga^i(\lla_{n-i} +b) +\gga^i\cdot 0}
 \,,
 }
where $\lla_n:=0$.  These elements of the group ring
$\overline{\CR}[A]$ are covariant under $A$ 
%in a way analogous to  \ainv,
%
 \eqn\ainva{
\ochi^{(b)}_{ax} =a\,\ochi^{(b)}_x \,, \qquad \forall a\in A \,.
}
We recall the expressions for the generalised
characters for the $n=2,3$ cases for which
$\tC=A\,t_{-P_+}$
 and $\tC= A\,t_{-P_+}\cup
A\,w_0\,t_{-P_+}$  respectively \FGPb\
\eqn\examp{
\eqalign{
\chi_x & = \ochi^{(0)}_x={\rm det}(\overline x)\, 
\big(\ochi_{\nu}+  \ga\ \ochi_{\nu- {\za\over 2}} \big)\,, \qquad
\qquad \quad n=2\,,\cr
\chi_x & = \ochi^{(0)}_x + \chi_{w_0} \ \ochi^{(\theta)}_x\,,
\qquad \chi_{w_0}= 2+w_0+w_1+w_2\,, \quad  n=3\,.
}}
The square of the $A$-invariant combination $F:=
w_0+w_1+w_2=aFa^{-1}$ in \examp\ lies in $\overline{\CR}[A]$.
%this implies that the ring generated by the formal characters in
%\examp\ can be interpreted as a quadratic extension of the
%classical  $sl(3)$ character ring  $\overline{\CR}$.

We now turn to the $sl(4)$ case. The fundamental chamber
$\tC$ is alternatively represented as $\tC\equiv\UU\,t_{-P_+}$, where
\eqn\us{
\UU=\{A\,, A\,w_0\,, A\,w_{10}\,, A\,w_{30}\,, A\,w_{310}\,,
A\,w_{2310} \}\,
}
is a subset of $\tW$; its
 projection $\overline{\UU}$ onto the subgroup $\bW$ gives the
right cosets of  $\bar{A}$.   
The group  $A$
defines  an automorphism of $W$, $w_{\za}\rightarrow \zg \,
w_{\za} \zg^{-1}=w_{\zg(\za)}\,,$ $\za\in \Pi$, with
$\zg(\za_j)=\za_{j+1}$ for $j=0,1,2,\dots n-1,$ identifying
$\za_{n}\equiv\za_0$. Using this we define  $A$ - invariants
 in the group algebra of $W$,
$F_y=F_{aya^{-1}}=a F_y a^{-1}$,  $\forall a\in A$,
%ring $\IZ[W]$
%
\eqn\df{
F_{rst...}\equiv F_{w_{rst...}}:={1\over l_{w_{rst...}}}\,
 \sum_{a\in A}aw_{rst...}a^{-1}\,,
\quad w_{rst...}\in W\,,
}
where $l_w$ takes the value $1$ or $2$ if the sum over $A$ contains
$4$ or $2$ different terms, respectively; e.g., 
$F_0=w_0+w_1+w_2+w_3\,,$ $\, F_{13}=w_{13}+w_{02}=F_{20}$.  As it
is clear from \tlg, the terms in a given $F_y$ have their $\iota$
images
in  an  orbit of the cyclic subgroup $\bar{A}$  of $\bW$.
%Clearly $F_y=F_{aya^{-1}}$ for $\forall a\in A$ while for the
%product of two such elements we have, with the multiplication
%inherited from  that in $\IZ[W]$,
%\eqn\prf{ F_x\ F_y = \sum_{a\in A}\, F_{xaya^{-1}}=\sum_{a\in A}\,
% F_{axa^{-1}y}\,.}
%
In general $F_x\ F_y \not = F_y\ F_x$, but e.g., the three
elements $Y_0:=F_0\,, Y_{30}:=F_{30}+F_{13}\,,
Y_{10}:=F_{10}+F_{13}\,, $ commute between themselves.

We shall introduce now a finite set of formal characters
$\cc_y\,, y\in \CC:= \tC\cap W\,,$ as in $\res$, for all of which
we will adopt the definition  $\mt$.  In employing the map \mi\
and comparing with the standard $sl(4)$ weight diagrams one can
use the recursive  formula for the multiplicity of a weight $\mu$
(see, e.g. \Zh{})
\eqn\fre{
\overline{m}_\mu =\;
-\sum_{\overline{w}\in\overline{W};\,\overline{w}\not = {\bf 1}} 
{\rm det}(\overline{w})\  \overline{m}_{\mu +\rho - \overline{w}(\rho ) }\,,
}
with the weights in the r.h.s.  strictly greater than $\mu$.
Using \fre\ we have  for $y\in \CC$ and of length $l(y) \le 3$
\eqna\Ia
$$\eqalignno{
\cc_{w_0}& 
=3+F_0 \equiv 3 + Y_0\,, \qquad \qquad \qquad\qquad \qquad \qquad
\iota(w_0)=(1,0,1)\,,\cr
\cc_{w_{10}} & = 3 + 2F_0 + F_{13}  + F_{10}\equiv 3+2Y_0+Y_{10}
\,,\ \ \qquad 
\iota(w_{10})=(0,1,2)\,,\cr
\cc_{w_{30}}& = 3 + 2F_0 + F_{13}  + F_{30}\equiv 3+2Y_0+Y_{30}
\,, \ \  \qquad 
\iota(w_{30})=(2,1,0)\,, \cr
\cc_{w_{130}} & = 7 + 5F_0 + 4 F_{13} + 2F_{30} + 2F_{10} +
(F_{121} + F_{130} + F_{213}) \, &\Ia{} \cr
&= : 7 + 5Y_0  + 2Y_{10} + 2Y_{30} + Y_{130}\,,
\qquad \qquad \qquad \iota(w_{130})=(1,2,1)\,,\cr
\cc_{w_{230}} & 
= 1 + F_0 + F_{13} +
 F_{30} + F_{230} 
\equiv 1  + Y_0 + Y_{30}+ \zg\,\bc_{\fw_1}\,, \quad
\iota(w_{230})=(4,0,0)\,, \cr
\cc_{w_{210}}& 
= 1 + F_0 + F_{13} + F_{10} +
 F_{210} 
\equiv 1 + Y_0 + Y_{10} + \zg^3\bc_{\fw_3}\,, \ \
\iota(w_{210})=(0,0,4)\,,\cr
}$$
of dimension $7,17,17,63, 15,15,$  respectively
(here $(a_1,a_2,a_3)=\sum_i\, a_i\bL_i$).  Being linear
combinations of $A$-invariant elements they commute with any
$a\in A$.  To each of these characters we associate a weight
diagram which can be   identified with a finite subset of the
Cayley graph of $W$ (see \FP\ for a schematic drawing of the
latter).  In agreement with  \wsa\ one obtains by a direct
computation
\eqn\fr{
\eqalign{
\cc_{w_0}\ \cc_{w_0} &= 1 + 2\, \cc_{w_0} + \cc_{w_{10}} + 
     \cc_{w_{30}}\cr 
\cc_{w_0}\ \cc_{w_{10}} &= \cc_{w_0} + 2\,\cc_{w_{10}}  + 
     \cc_{w_{130}}    + \cc_{w_{210}}\cr
\cc_{w_0}\ \cc_{w_{30}} &= \cc_{w_0} + 2\,\cc_{w_{30}} + \cc_{w_{130}} + 
     \cc_{w_{230}}\,,
}}
which serve as algebraic relations restricting the set of characters
\eqn\bse{
{\cal F}=\{
 \cc_{w_0}\,,\cc_{w_{10}}\,,\cc_{w_{30}}\,,
\cc_{w_{210}}\,,\cc_{w_{230}}\}\,.
}
Further fusions recover the characters of length $4$, in particular
the  character $\chi_{w_{2130}}$  of dim 177, with $\,
\iota(w_{2130})=(2,2,2)$, 
\eqna\vio{
$$\eqalignno{
\chi_{w_{2130}} &=
\cc_{w_{10}}\ \cc_{w_{30}} -\Big(1+2\, \cc_{w_{0}}+\cc_{w_{10}}+ \cc_{w_{30}}+
\cc_{w_{130}}\Big)\cr
&
= 11 + 9F_0 + 8\, F_{13} + 4 F_{10} + 4 F_{30} + 3F_{121} +
3 F_{130} + 3 F_{213}+ 2F_{230} + 2F_{210}\cr
& + (F_{10} +  F_{30}+ F_{1213} +  F_{1232} +
F_{1321} +F_{2321} + F_{0213}  + F_{2130}) &\vio{} \cr
&=: 11 +
9Y_0  + 4 Y_{1} + 4 Y_{3} +3 Y_{130} +   2  \zg\bc_{\fw_1}+ 2
\zg^3\bc_{\fw_3} +Y_{2130}\,.
}$$
This is the  simplest example in which formula \mt\
fails. The  expression obtained by  fusion
corresponds to a weight diagram that is a subset of the one
determined by
\mt\ and \fre. It can be summarised by the rules: i) delete all elements
longer than the highest weight element; ii) decrease the
multiplicity of $w_{ijk...}$, determined from \mt, by the
complement to 4 of the number of different elementary reflections
appearing in $w_{ijk...}$.  E.g. for $w_{2321}$ the multiplicity
\mt\ is decreased by $1$ since three of the four reflections
appear, while for $w_{0213}$ it is left unchanged; the
multiplicity of the identity is decreased by $4$.

One obtains by a direct computation   the relations
\eqn\algr{
\eqalign{
Y_0^2 &= 4 +Y_{10} +Y_{30}\cr
Y_{10}^2 &= 2+3Y_{30}-Y_{10} +Y_0\ \gamma^3 \ochi_{\bL_3}
+\gamma^2 \ochi_{\bL_2}\cr 
Y_{30}^2 &= 2+3Y_{10}-Y_{30} +Y_0\ \gamma \ochi_{\bL_1}+\gamma^2
\ochi_{\bL_2}\cr 
Y_0 \ Y_{10} & = 2Y_0 +Y_{130}+ \gamma^3 \ochi_{\bL_3} \cr
Y_0 \ Y_{30} & = 2Y_0 +Y_{130}+ \gamma \ochi_{\bL_1}\cr
Y_{10}\ Y_{30} & = 6 +Y_{2130}\,,
}}
which imply that all five elements  $Y_0, Y_{10}, Y_{30},
Y_{130}, Y_{2130}$ commute.  Because of the first three
equalities the product of any two of these elements is expressed
as a linear combination of the same  elements plus the identity,
with coefficients in the ring $\overline{\CR}[A]$.  Alternatively
the relations \algr\  can be rewritten as
\eqna\poly
$$\eqalignno{
&Y_{10}+Y_{30}=Y_0^2-4=:P_2(Y_0)\cr
&2 Y_{130}=Y_0^3- 8Y_0 -(\gamma^3 \ochi_{\bL_3}+\gamma
\ochi_{\bL_1})= :P_3(Y_0) &\poly{}\cr 
&2Y_{2130}= Y_0^4 -10 Y_0^2 -Y_0(\gamma^3 \ochi_{\bL_3}+\gamma
\ochi_{\bL_1})+8 -2 \gamma^2 \ochi_{\bL_2}=:P_4(Y_0) \cr 
 (Y_{10}&-Y_{30})(\gamma^3 \ochi_{\bL_3}-\gamma\ochi_{\bL_1} )
=-Y_0^5+ 12Y_0^3
+2Y_0^2(\gamma^3 \ochi_{\bL_3}+\gamma
\ochi_{\bL_1})\cr
&  + 4 Y_0 (\gamma^2 \ochi_{\bL_2}-6)=:P_5(Y_0)\,,
}$$ 
where $P_k(Y_0)$  are $k$-order polynomials of $Y_0$, and
furthermore, $Y_0$ satisfies a $6$-order polynomial equation,
\eqn\pol{
Y_0^6 -12 Y_0^4 - 2 c\, Y_0^3 - 4 Y_0^2(\gamma^2 \ochi_{\bL_2}-6)  +
c^2-4(1+\ochi_{\theta})=0 
}
where $c=\gamma^3 \ochi_{\bL_3}+\gamma \ochi_{\bL_1}$. The relations
\algr\ suggest that the formal characters we look for
are given as  linear combinations of the six elements $Y_{g}\,,\,
g\in \UU_{\un}:=\{\un\,,\, w_0\,,\,w_{10}\,,\,w_{30}\,,\,
w_{130}\,,\, w_{2130}\} 
\subset\CU$, with coefficients in the ring $\overline{\CR}[A]$.
We define
\eqn\gchar{
\eqalign{
\chi_x: 
 &=\sum_{g\in \UU_{\un}}\, \chi_g\ \sum_b\, c_{g,b}\ \ochi_x^{(b)}=
 (\ochi^{(0)}_x -\ochi^{(\theta+ \za_2)}_x)
  +\chi_{_{0}}\ (\ochi^{(2 \theta)}_x+\ochi^{(\theta)}_x)\cr
&+ \chi_{_{10}}\ \ochi^{(2\theta- \za_1)}_x + \chi_{_{30}}\ 
\ochi^{(2\theta- \za_3)}_x +
\chi_{_{130}}\ \ochi^{(\theta+ \za_2)}_x + 
\chi_{_{2130}}\ \ochi^{(\theta)}_x \,. 
 }}
Choosing $x=g$  for $g\in \CU_{\un}$ \gchar\ reproduces the
formulae  for the   characters in \Ia{}, \vio{}.  The values of
the shifts $b$ are recovered from each of these five basic
characters demanding that the first classical character of the
quadruplet \quadr\ is an identity. This gives $b_g=0$ for the
identity $g=\un$, while $b_g= - \bar{g}\cdot(-\theta)$
for  $g=\bar{g}\, t_{-\theta}$, i.e.,
 $b_g= 2\theta\,,\, 2\theta- \za_1\,,\, 2\theta- \za_3\,,\,
 \theta + \za_2\,,\, \theta$, respectively.
In these checks one has to use repeatedly the symmetry \class\ of the
classical characters to cancel abundant terms. Similarly
one finds $\chi_a=a$ for any $a\in A$. 
The proposed expression \gchar\ is justified by the

\LEMMA The following Pieri-type formulae hold true for the
characters defined in \gchar\ and any $f\in {\cal F}$:
\eqn\pieri{
\chi_f\ \chi_x = \sum_{w\in \CG_f}\ \chi_{wx}\,.
}
The first two   relations  are proved  by a direct but
tedious computation comparing the products in the l.h.s. with the
r.h.s. of \pieri.  It is based on the polynomial relations \algr\
satisfied by the invariants $Y_g$.  
%It is sufficient to check the first two relations as 
 One has also to use the
classical characters multiplication tables of the fundamental
characters $\ochi_{\bL_i}\,, i=1,2,3$ and $\ochi_{\theta}$, which
extend to the multiplication rules
\eqn\cla{
\ochi_{\lambda}\ \chi_x = \sum_{\mu\in \Gamma_{\lambda} }\ \chi_{t_{-\mu}\,
x}\,. 
}
The third relation  is recovered from the second by
the symmetry $w_1 \leftrightarrow w_3$.
The proof 
%of the Pieri formulae
 for the last
two characters $\chi_{w_{230}}, \chi_{w_{210}}$ uses
the fact that they are expressed in terms
of the first three characters in \bse\ (cf. \Ia{})
 and the fundamental classical characters
$\ochi_{\bL_i}\,, i=1,2.$\endPROOF

Formulae \pieri\ and \gchar\  hold for generic $x$, sufficiently
far from the walls of the chamber $\tC$, otherwise cancellations
occur, as e.g., in the examples
\fr. Using \ainva\ the ``fundamental'' 
multiplication  formulae \pieri\ are extended to include
the "simple currents"    corresponding to
 the elements $a=\cc_a\,$ of $A$ 
\eqn\ainv{
\cc_a\ \cc_x
=\cc_{ax}\,.
}
Having the explicit formula \gchar\ one can compute the
multiplicities in \res.  
%The property \ainva\ extends to \ainv\
%for the generalised characters \gchar.  
Furthermore we recall
that the following proposition was proved in \FP\ under the
assumption that \pieri\ holds true:
\PROP {\it  For any $y\in \CC$ there is a formal character $\cc_y$
obtained recursively, using \pieri, as a polynomial of 
the commuting ``fundamental'' characters in
%of the set ${\cal F}$ 
 \bse.}

\medskip
\noindent
Using \ainv\ the Proposition is extended to $y\in \tC$. 
 Combined with associativity it furthermore
allows to extend \pieri, \ainv\ to the multiplication of arbitrary two
characters as in the first equality in \ws, and hence to confirm
 formula \wsa\ for the fusion multiplicities.  The proof is a
straightforward generalisation of the second proof of  Lemma 4.5
in \FGPb. What however remains to be proved in general is the
non-negativity of the multiplicities $N_{x,y}^z$ in \wsa; so far
we have checked it on numerous examples.

\noindent
\newsec{Discussion}   

\vskip 0.1cm
Extending the results of \FGPb\ we have found a consistent
$\widehat{sl}(4)_k$ fusion ring generated by the formal
characters \gchar.  To interpret it as the fusion ring of the
related quasi-rational WZNW field theory one has to show that the
(shifted) generalised weight diagrams of the generating
characters in \pieri, \Ia{}\ are consistent with the solution of
the equations expressing the  decoupling of the corresponding
Verma module singular vectors.  As in \FGPa,\GPW\ we can use the
standard functional realisation of the representations of
$sl(4)$, in which the generators are represented by differential
operators in $6$ variables, see e.g., \Zh. The resulting systems
of partial differential equations are however  rather involved
and we have checked the simplest of them, corresponding to the
``fundamental'' representation labelled by $w_{230}=\gamma\,
t_{-\bL_1}$: the 15 points of the generalised weight diagram  in
\Ia{}\ are confirmed. We have also partially checked the
multiplication rule of the generator $\chi_{w_0}$, choosing a
particular target representation for which the system of
equations simplifies: once again the generalised weight diagram
in \Ia{}\ consisting of 4 points of multiplicity $1$ and one
point of multiplicity $3$ is confirmed.
 
In the rational case  $k+4=4/p$ ($p$ - odd) the roots of the
equation \pol\ determine $Y_0$ (and hence all five generators
expressed by the polynomials $P_k(Y_0)$)  in terms of the
integrable representations fusion ring  characters
$\ochi_{\lambda}^{(p)}(\mu)$ at  level $p$.  In principle this 
should allow  the ``quantisation'' of the general characters in
\gchar, as it has been achieved in the sl(3) case in \FGPb.
Similarly we can use \pol\ in order to  evaluate the formal
characters \gchar\ on the Cartan subgroup of $SU(4)$. 
%{\it but need to give meaning to $a(h)\in A$?.} 
In particular their values
at the identity give the dimensions, alternatively obtained by
sending all $z\to 1$ in \res. 

The method is expected to apply algorithmically to any $n$ starting
with the analogues of the set \us\ and determining the coefficients
$c_{g,b}$ in the analogue of \gchar\ from the ``fundamental''
fusions generalising \pieri. The $6$-order polynomial will be replaced
by a $(n-1)!$-order polynomial.
The non-trivial problem that remains is to find a universal
formula for the weight multiplicities in \ws, extending \mt,
which in particular would allow to prove the non-negativity of
the structure constants in \wsa.

\vskip 1cm
\noindent
{\bf Acknowledgments}
\vskip 0.1cm
P.F. acknowledges the 
 financial 
support of the 
%Italian Ministry of
University of Trieste.
%, Scientific Research and Technology (MUSRT). 
 V.B.P. acknowledges the  support and hospitality of
INFN, Sezione di Trieste and ICTP, Trieste and  the hospitality
of the School of Computing and Mathematics, University of
Northumbria, Newcastle, UK.

\vskip 1cm

\noindent
\listrefs
\end

==============================

\lref\AY{Awata H. and  Yamada Y., 
       Fusion rules for the fractional level $\widehat{sl}(2)$ 
algebra,         Mod. Phys. Lett. {\bf A7}, 1185--1195 (1992).}

\lref\FMa{Feigin, B.L. and Malikov, F.G. 
        Fusion algebra at a rational level and cohomology of 
           nilpotent subalgebras of $\widehat{sl}(2)$,
      Lett. Math. Phys. {\bf 31}, 315--325 (1994).}
\lref\FMb{Feigin, B.L. and Malikov, F.G. 
 Modular functor and representation 
        theory of $\hat{sl(2)}$ at a rational level, in
         {\it Operads}: Proceedings of Renaissance Conferences,  
         Cont. Math. {\bf 202}, p. 357, J.-L. Loday, J.D. Stasheff and 
        A.A. Voronov, eds. (AMS, Providence, Rhode Island 1997),
 q-alg/9511011.}

\lref\MFF{Malikov, F.G., Feigin, B.L., and Fuks, D.B.,

     	  Funct.\ Anal.\ Pril.\ {\bf 20, 2} (1987) 25.}

\lref\FGP{Furlan P., Ganchev A.Ch. and Petkova V.B., 
             Quantum groups and fusion rule multiplicities,
              Nucl. Phys. {\bf B343}, 205--227 (1990).} 

\lref\FGPa{Furlan P., Ganchev A.Ch. and Petkova V.B.,
        Fusion rules for admissible representations of affine 
        algebras:           the case of $A_2^{(1)}$, 
         Nucl. Phys. {\bf B518} [PM], 645--668 (1998), hep-th/9709103.}
 
\lref\FGPb{Furlan P., Ganchev A.Ch. and Petkova V.B.,
         An extension of the character ring of $sl(3)$ and its quantisation,
		Comm. Math. Phys. {\bf 202 } 701--733 (1999), math.QA/9807106.}
		
\lref\GPW{Ganchev A.Ch., Petkova V.B. and Watts G.M.T., A note
on decoupling conditions for generic level $\widehat{sl}(3)_k $
and fusion rules, 
Nucl. Phys. {\bf B571} [PM] 457--478 (2000),
 hep-th/9906139.}
 
\lref\Hump{Humphreys J.M., {\it Reflection Groups and
           Coxeter Groups}, (Cambridge University Press, 1990).} 

\lref\K{Kac V.G., {\it Infinite-dimensional Lie Algebras}, third
        edition, (Cambridge University Press, 1990).}

\lref\KW{Kac V.G. and  Wakimoto M.,  
      Modular invariant representations of infinite-dimensional 
      Lie algebras and superalgebras,
    Proc. Natl. Sci. USA {\bf 85}, 4956--4960 (1988)   \semi 
 Kac V.G. and  Wakimoto M.,
         Classification of modular invariant 
                 representations of affine algebras,
               Adv. Ser. Math. Phys. vol {\bf 7}, pp. 138--177. 
(World Scientific, Singapore, 1989)         \semi
 Kac V.G. and  Wakimoto M., 
        Branching functions for winding subalgebras and tensor 
products, 
          Acta Applicandae Math. {\bf 21}, 3--39 (1990).}

\lref\Zh{Zhelobenko D.P., {\it Compact Lie Groups and their Representations}, 
(AMS, Providence, Rhode Island, 1973).}

\lref\MW{Walton M., 
       Fusion rules in Wess-Zumino-Witten models, 
           Nucl. Phys. {\bf B340}, 777--790 (1990). }

\lref\FP{Furlan P., Ganchev A.Ch. and Petkova V.B.,
        Fusion rules for admissible representations of affine 
        algebras:           the case of $A_2^{(1)}$, 
         Nucl. Phys. {\bf B518} [PM], 645--668 (1998)\semi
 Furlan P., Ganchev A.Ch. and Petkova V.B.,
         An extension of the character ring of $sl(3)$ and its quantisation.
		Comm. Math. Phys. {\bf 202 } 701--733 (1999). }

\lref\Hump{Humphreys J.M., {\it Reflection Groups and
           Coxeter Groups}, (Cambridge University Press, 1990).} 

\lref\K{Kac V.G., {\it Infinite-dimensional Lie Algebras}, third
        edition, (Cambridge University Press, 1990).}

\lref\KW{Kac V.G. and  Wakimoto M.,  
      Modular invariant representations of infinite-dimensional 
      Lie algebras and superalgebras,
    Proc. Natl. Sci. USA {\bf 85}, 4956--4960 (1988)   \semi 
 Kac V.G. and  Wakimoto M.,
         Classification of modular invariant 
                 representations of affine algebras,
               Adv. Ser. Math. Phys. vol {\bf 7}, pp. 138--177. 
(World Scientific, Singapore, 1989)         \semi
 Kac V.G. and  Wakimoto M., 
        Branching functions for winding subalgebras and tensor 
products, 
          Acta Applicandae Math. {\bf 21}, 3--39 (1990).}

\lref\Zh{Zhelobenko D.P., Compact Lie Groups and their Representations, 
(AMS, Providence, Rhode Island, 1973).}

\lref\MW{Walton M., 
       Fusion rules in Wess-Zumino-Witten models, 
           Nucl. Phys. {\bf B340}, 777--790 (1990). }

\lref\FP{Furlan P. and Petkova V.B.,  
Fusion Rings Related to Affine Weyl Groups,
in: Proc.
%eedings 
of the  International Workshop 
{\it Lie Theory and Its Applications in Physics III}, 
(Clausthal, 1999); eds. H.-D. Doebner et al, 
(World Sci, Singapore, 2000, ISBN 981-02-4421-5) pp. 237-249, 
hep-th/0007219.}

\lref\ZZ{ Teschner, J.
 Liouville theory revisited,  hep-th/0104158.}
 
\lref\GKS{Giveon, A,  Kutasov, D. and  Schwimmer, A.,
     Comments on D-branes in AdS$_3$, hep-th/0106005.  }
	 
\lref\admFGP{Furlan P., Ganchev A.Ch. and Petkova V.B.,
$A_1^{(1)}$ admissible representations --fusion transformations
and local correlators, Nucl. Phys. {\bf B491}  no. 3 [PM] 635-658
(1997), hep-th/9608018.}

\lref\PRY{ J.L. Petersen, J. Rasmussen and M. Yu,  Fusion, Crossing
      and monodromy in conformal field theory based on $sl(2)$ current
      algebra with fractional level,  Nucl. Phys. {\bf B 481} 577-624 (1996),
       hep-th/9607129.}

%%%%%%%%%%%%%%%%%%%%%%%%%%%%%%%%%%%%%%%%%%%%%%%%%%%%%%%%%%%%%%%%%%%%